\documentclass[letterpaper,12pt]{article}   
\usepackage{graphicx}
\usepackage{amsmath}
\usepackage{epstopdf}

\begin{document}

\title{A Cook Book of Structure Functions}

\author{Agla\'e Kellerer\footnote{a.n.c.kellerer@durham.ac.uk, Physics Dept., Durham University, Durham, DH1 3LE}}
\maketitle

{\bf Abstract\/} The structure function is a useful quantity to characterize wavefront distortions. 
We derive  expressions for the structure functions of the averaged wavefront phase and slopes. The expressions are valid within the inertial range of atmospheric turbulence, and are meant to serve as engineering formulae when reconstructing profiles of the atmospheric turbulence, specifically in the context of atmospheric profiling instruments (e.g. SLODAR and S-DIMM+) and multi-conjugate adaptive optical systems.

\section{Structure function}

Kolmogorov's theory of turbulence provides the structure function of phases in a wavefront. It equals the expected mean squared difference of the phases,  $f$ and $f'$, at two points, $(x,y)$ and $(x',y')$, separated by distance $s$:
\begin{align} 
SF(s) = \{ (f - f')^2 \}   = k \, \left(\frac{s}{r_0}\right)^{5/3}
\label{eq:1}
\end{align} 
with $k = 6.88$. $r_0$ is the Fried parameter: in a circular region with a diameter equal to the Fried parameter, $r_0$, the phase variance is roughly equal to 1 (see for example Roddier\,\cite{Roddier}).
The symbol $\{ \cdot \}$  stands for the statistical expectation.

$SF(s)$ relates to the point function, $f(x,y)$, and to a coherent wavefront. In actual adaptive-optics (AO) conditions one deals with averaged phases or phase-slopes, since measurements utilize sensors of finite size. Accordingly, the structure functions  need to be adjusted to apply to the averaged phases or slopes.

Expressions for the adjusted structure functions have been derived by a number of authors, see Tokovinin\,\cite{Tokovinin} and references therein. These expressions were derived for use with the Differential Motion Monitor (DIMM)\,\cite{DIMM} and are valid for separations larger than the averaging diameter, $s>d$. 

In order to reconstruct profiles of the atmospheric turbulence, it is useful to extend the expressions to separations that are smaller than the averaging diameter, see for example Scharmer \& van Werkhoven\,\cite{Scharmer} and Kellerer et al.\,\cite{SDIMM}.
Here we derive expressions valid within the inertial range of atmospheric turbulence, i.e. the range of separations where Eq.\,\ref{eq:1} correctly approximates the structure function of the phase. 

\section{Structure function of the degraded phase}
 
Let $g$ be the phase averaged over a (circular or square) region $A$ centered at $(x,y)$:
\begin{align} 
g(x,y)  =  < f(x',y') >_A
\label{eq:5}
\end{align} 

The notations adopted in the following text are explained in Appendix\,\ref{app:1}. 

As with the undistorted phase, $f$, the variance of $g$ cannot be given without specifying a reference region. This is not the case for the structure function. 

Let $A$ be a square of size $d\times d$ or a circle of diameter $d$.
For large distances $s$ the structure function, $SF(s;d)$, of $g$ converges to the structure function, $SF(s)$, of $f$.  How the functions differ at smaller distances $s$ needs to be investigated:
\begin{align} 
SF (s;d)  &= \{ (g(x,y) - g(x+s,y))^2 \}   \\
	&=   \{ (<  f(x,y) >_A - < f(x+s,y) >_A)^2 \} \\
	&=  2 \{ <  ( f(x,y) >_A^2 - < f(x,y) >_A \, < f(x+s, y) >_A \}  \\
	&=   2 \{ <  f(x,y) f(x',y') - f(x,y)\,f(x'+s, y' ) >_A \} \\
	&=   \{ <  ( f(x,y) - f(x'+s, y' ))^2 -  (f(x,y) - f(x',y' ))^2 ) >_A \}\\
	&=    k  \, <  (u'^{5/3} - u^{5/3}) >_A /r_0^{5/3}
\end{align} 
with:    $u = ((x-x')^2 + (y-y')^2)^{0.5} $ and $u'= ((x-x'+ s)^2 + (y-y')^2)^{0.5}$.  	 
In standard notation this result reads:
\begin{align} 
SF (s;d)     =     k \int_A  \int_A   (u'^{5/3} - u^{5/3})  \, dx \, dy \, dx' \, dy'/ S^2
\end{align} 
$S$: area of $A$ ($d^2$ for a square, $\pi\,d^2/4$ for a circle). 

To put this into words: Let $A'$ be the averaging region $A$ shifted by distance $s$ in the direction $x$. The degraded structure function is then equal to the average structure function between a point in $A$ and a point in $A'$, minus the average structure function between two points within $A$.  For $s >>r$  the degraded structure function, $SF (s;d)$, converges to $SF (s)$. 

Since both $SF(s;d)$ and the unmodified structure function $SF(s) = k\,(s/r_0)^{5/3}$ increase steeply with $s$ the comparison of the structure function can best be made in terms of the ratio of the two functions. This ratio can be termed the reduction factor, \begin{align} 
RF(s;d)=\frac{SF(s;d)}{k\,(s/r_0)^{5/3}}
\end{align} 
It specifies -- for two points separated by distance $s$ -- the reduction of the mean squared phase difference due to the phase averaging over diameter $d$.

\begin{figure*}
\centering
\includegraphics[width=.45\textwidth]{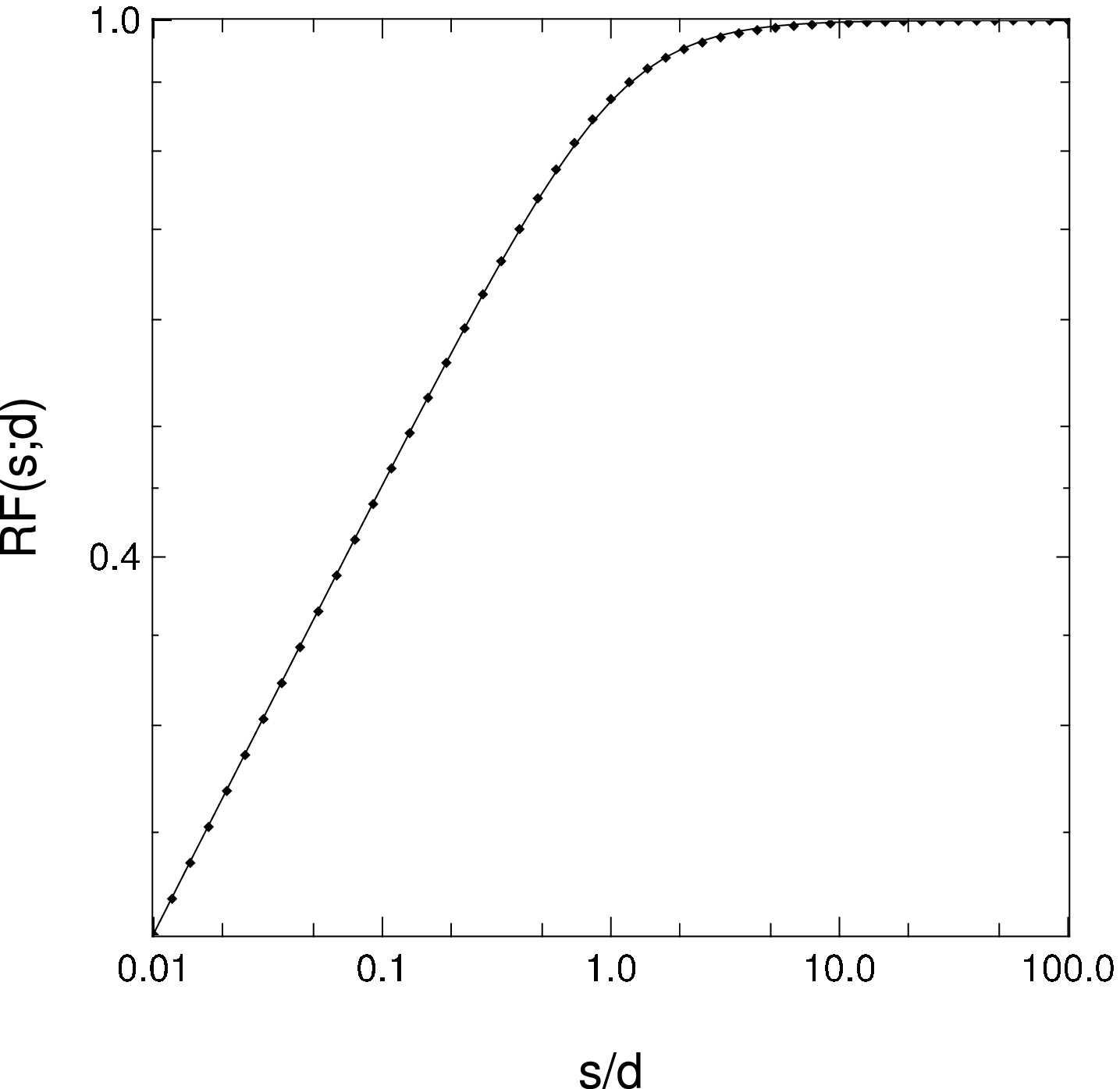}
\includegraphics[width=.45\textwidth]{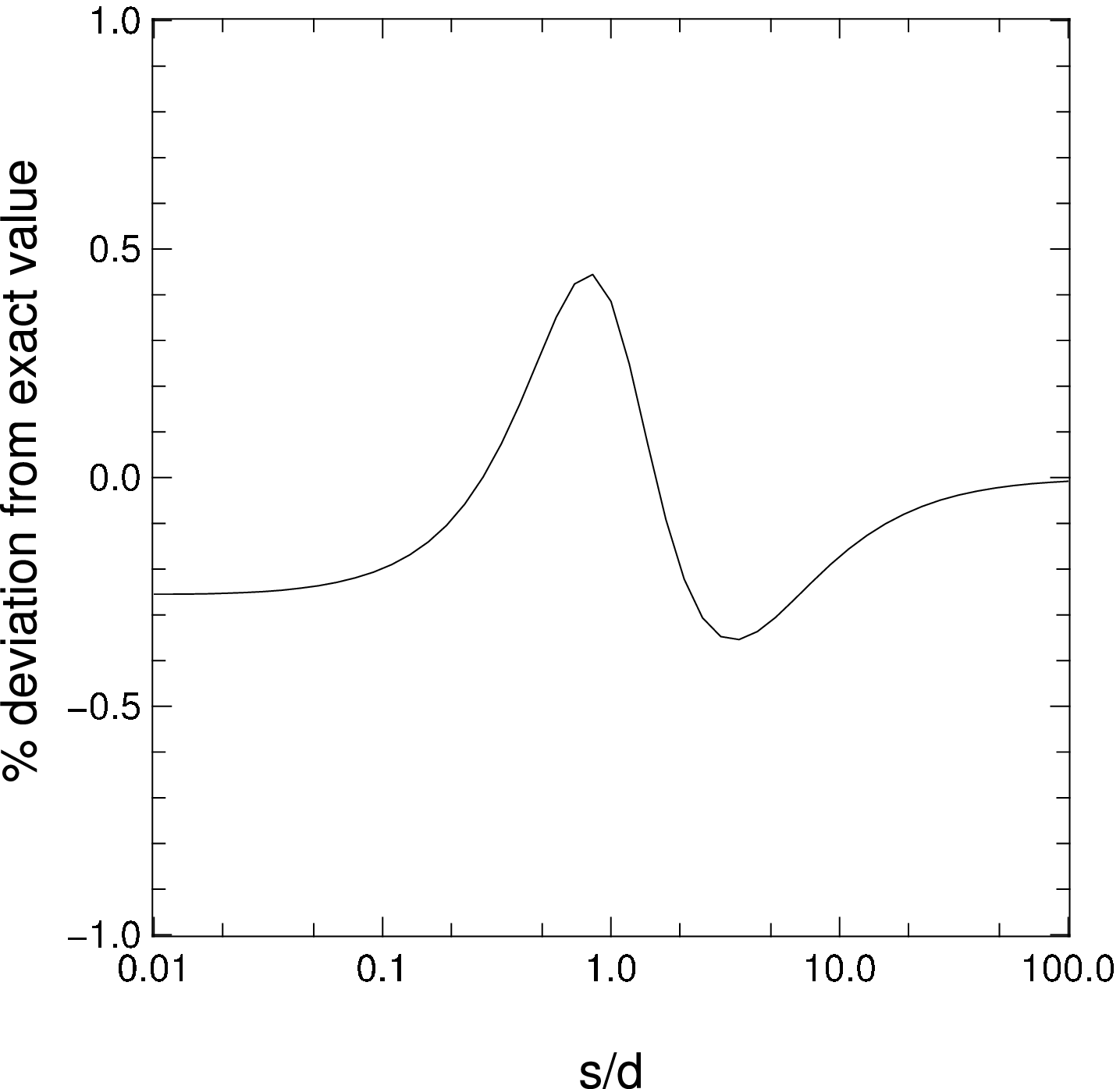}
\caption{Left panel: Reduction function obtained from $10^6$ random point-pairs. The standard deviation is too small to be identified. Dots: Numerical evaluation, line: analytical approximation in terms of Eq.\,\ref{eq:app}. Right panel: Deviation of the analytical approximation  to the numerical values. }
\label{fig:RF}
\end{figure*}

\paragraph{Approximate formulae} The numerical evaluation is shown on Fig.\,\ref{fig:RF}, it is obtained from $10^6$ point pairs chosen randomly within a circle of diameter 1. The following analytical function approximates the result with less than 0.5\% deviation over the range $s/d= 0.01 - 100$ (see right panel of Fig.\,\ref{fig:RF}):  
\begin{align} 
RF(s;d) &\sim (1+1.14\,(s/d)^{-5.5/3})^{-1/5.5}\\
SF(s;d) &\sim k\,\left(\frac{s}{r_0}\right)^{5/3}\,(1+1.14\,(s/d)^{-5.5/3})^{-1/5.5}
\label{eq:app}
\end{align} 

\section{Structure function of the wavefront slope}

\paragraph{Line-averaged slope} The simplest approximation to the slope is the phase difference between two points divided by their distance, $d$: 
\begin{align} 
s_x (x,y) &= \frac{\lambda}{2\pi} \, (f(x+d, y) - f(x,y))/d \\
s_y (x,y) &= \frac{\lambda}{2\pi} \, (f(x, y+d) - f(x,y))/d
\end{align} 

\paragraph{Mean slope} A related  notion  is the slope averaged not over a line element, $d$, but over a reference region, $A$. It equals the derivative of the degraded phase $g(x,y)$ (see Eq.\,\ref{eq:5}): 
\begin{align} 
s_x  &= <  {\rm d\/} f(x',y') /{\rm d\/}x  >_A = {\rm d\/} g(x,y)/{\rm d\/}x  \\
s_y  & = <  {\rm d\/} f(x',y') /{\rm d\/}y  >_A= {\rm d\/} g(x,y)/{\rm d\/}y
\end{align}

\paragraph{Least-square slope} A further concept is the slope of the least-square linear approximation to the wavefront over the reference region. It is particularly relevant because it is a close approximation of the quantity measured by a SH-sensor\,\cite{Tokovinin2007}.

To avoid an excess of symbols, the same letters, $s_x, s_y$, are used for the three different choices; the type of average needs to be recognized from the context.

\subsection{Average gradient over a line segment }
The simplest approximation to the slope is the phase difference between two points a distance $d$ apart:
\begin{align} 
s_x(x,y) &= \frac{\lambda}{2\pi} \, (f(x+d, y) - f(x,y))/d \\
s_y(x,y) &= \frac{\lambda}{2\pi} \, (f(x, y+d) - f(x,y))/d
\end{align} 

In the equation for the structure function of $s_x$ only the variable $x$ appears, which makes it convenient to use the simpler notation $f_x$ for $f(x,y)$, etc.
In this simplified notation the structure function of the slope, $s_x$, in the direction of the distance $x$, reads:
\begin{align} 
SF_x (s;d) &= \{ (s_x (s,0) - s_x(0,0))^2 \} \\
&= \left(\frac{\lambda}{2\pi}\right)^2 \, \{ (f_{s+d} - f_s - f_d +f_0)^2 \} / d^2 \\
&= \left(\frac{\lambda}{2\pi}\right)^2  \, \Big(\{ (f_d-f_0)^2 \}  + \{(f_d-f_{s+d})^2\} - \{(f_d-f_s)^2\}   \\
 &- \{ (f_0-f_{s+d})^2 \}  + \{(f_0-f_s)^2\} - \{(f_{s+d}-f_s)^2\} \Big)/ d^2 
\label{eq:7}
\end{align} 

\paragraph{The classical DIMM formulae:}
If $s>>d$, this simplifies to:
\begin{align} 
SF_x (s;d) &= \left(\frac{\lambda}{2\pi}\right)^2  \,  ( 2\,SF(d)/d^2 - SF''(s) ) \\
&= 2 k\,\left(\frac{\lambda}{2\pi}\right)^2 \, r_0^{-5/3}\,( d^{-1/3} - 5/ 9 \, s^{-1/3}) \\
&= 0.35\, \lambda^2\, r_0^{-5/3}\,d^{-1/3}\,(1-5/9\,(s/d)^{-1/3})  
\label{Sar_x}
\end{align} 
Similarly, for the transverse slopes:
 \begin{align} 
SF_y (s;d) &= 2\,\left(\frac{\lambda}{2\pi}\right)^2  \,( SF(d) + SF(s) - SF((s^2+d^2)^{0.5}) )/d^2 
\end{align} 
With 
 \begin{align} 
SF((s^2+d^2)^{0.5}) &= k\, (s^2+d^2)^{5/6} \,r_0^{-5/3}\\
&\sim k\,(s^{5/3} + 5/6\,s^{-1/3}\,d^2)\,r_0^{-5/3} = SF(s) +5/6k\,s^{-1/3}\,d^2\,r_0^{-5/3}
\end{align} 
one obtains the equivalent of Eq.\,\ref{Sar_x}:
 \begin{align} 
SF_y (s;d) &= 0.35 \, \lambda^2\, r_0^{-5/3}\,d^{-1/3}\,(1-5/6\,(s/d)^{-1/3})  
\label{Sar_y}
\end{align} 
The structure functions of the $x-$ and $y-$ slopes -- as used for DIMM -- are represented as blue dots on Figs.\,\ref{fig:SFx} and \ref{fig:SFy}. 

\paragraph{General relations:}
Eqs.\,\ref{Sar_x} and \ref{Sar_y} are limited to separations larger than the reference diameter: $s>>d$. 
They are the classical relations indicated by Sarazin and Roddier for the analysis of DIMM measurements. 
More general relations, valid for any value $s$, are:
\begin{align} 
SF_x (s;d) &= k\,\left(\frac{\lambda}{2\pi}\right)^2 \, r_0^{-5/3} \, (2 d^{5/3}  - | s - d |^{5/3} + 2 s^{5/3} - (s + d)^{5/3} ) /d^2\\
&= k\,\left(\frac{\lambda}{2\pi}\right)^2 \, r_0^{-5/3} \, d^{-1/3}\,(2 - | 1 - s/d |^{5/3} + 2 (s/d)^{5/3} - (1 + s/d)^{5/3} ) \\
SF_y (s;d) &= 2k\,\left(\frac{\lambda}{2\pi}\right)^2  \,r_0^{-5/3}  \, ( d^{5/3}  + s^{5/3}  - (s^2+d^2)^{5/6}  )/d^2  \\
&= 2k\,\left(\frac{\lambda}{2\pi}\right)^2  \, r_0^{-5/3}  \,  d^{-1/3}\, ( 1  + (s/d)^{5/3}  - (1+(s/d)^2)^{5/6}  ) 
\end{align} 
These two structure functions are indicated as blue lines on Figs.\,\ref{fig:SFx} and \ref{fig:SFy}.

\subsection{Average gradient over a reference region: G-tilt}

A related  notion -- which applies to a defocussed wavefront image --  is the slope averaged not over a line element, $s$, but over a reference region, $A$. It equals the partial derivative of $g(x,y)$ (see Eq.\,\ref{eq:5}): 
\begin{align} 
s_x  &= <  {\rm d\/} f(x',y') /{\rm d\/}x  >_A = {\rm d\/} g(x,y)/{\rm d\/}x  \\
s_y  & = <  {\rm d\/} f(x',y') /{\rm d\/}y  >_A= {\rm d\/} g(x,y)/{\rm d\/}y
\end{align} 
The relevant structure functions can be derived from the structure function of the averaged phase: 
\begin{align} 
SF_x (s;d)&= \left(\frac{\lambda}{2\pi}\right)^2  \, {\rm lim\/}_{\Delta\rightarrow 0} \left( 2\,SF(\Delta;d)/\Delta^2 - SF''(s;d) \right) \\
SF_y (s;d) &= 2\, \left(\frac{\lambda}{2\pi}\right)^2  \,{\rm lim\/}_{\Delta\rightarrow 0} \left( SF(\Delta;d) + SF(s;d) - SF((s^2+\Delta^2)^{0.5};d) \right)/\Delta^2 
\end{align} 

\paragraph{Approximate formulae}
We use the approximation for the structure function of the degraded function (see Eq.\,\ref{eq:app}) to derive the expressions for the structure function of the gradient over a circle:
\begin{align}
SF_x (s;d)&= \left(\frac{\lambda}{2\pi}\right)^2  \,k\,d^{-1/3}\,r_0^{-5/3}\, \Big( 2\,a^{-1/b} -\frac{1}{9}\,u^{-1/3}\,w^{-1/b} \,(10+a\,(7-b)\,u^{-b/3}\,w^{-1} +a^2\,(b+1)\,u^{-2b/3}\,w^{-2}  )\Big) \\
SF_y (s;d)&= 2\,\left(\frac{\lambda}{2\pi}\right)^2  \,k\,d^{-1/3}\,r_0^{-5/3}\, \Big( a^{-1/b} - \frac{1}{6} \,u^{-1/3}\,w^{-1/b}\,(5+a\,u^{-b/3}\,w^{-1}) \Big) 
\end{align}
where $a=1.14$, $b=5.5$, $u=s/d$ and $w=1+a\,u^{-b/3}$. This rewrites as, 
\begin{align} 
SF_x (s;d)&= \left(\frac{\lambda}{2\pi}\right)^2  \,k\,d^{-1/3}\,r_0^{-5/3}\, \Big( 1.95 
 -\frac{1}{9}\,u^{-1/3}\,w^{-1/5.5} \,(10+1.71\,u^{-5.5/3}\,w^{-1} +8.45\,u^{-11/3}\,w^{-2}  )\Big) \\
SF_y (s;d)&= 2\,\left(\frac{\lambda}{2\pi}\right)^2  \,k\,d^{-1/3}\,r_0^{-5/3}\, \Big( 0.98 - \frac{1}{6} \,u^{-1/3}\,w^{-1/5.5}\,(5+1.14\,u^{-5.5/3}\,w^{-1}) \Big)
\end{align}
These structure functions are represented as black dots on Figs.\,\ref{fig:SFx} and \ref{fig:SFy}.
The functions are in excellent agreement with the expressions given in Tokovinin\,\cite{Tokovinin} for $s>d$: see his Eq.\,7. Note that Tokovinin's equations are based on calculations by Conan et al.\,\cite{Conan}. 

\subsection{Least square fit to the wavefront: Z-tilt}

Another concept is the slope of the least-square linear approximation to the phase front over the reference region. This variable will be considered, because it equals the tilt measured by a Shack Hartmann (SH)-sensor:

Let $(0,0)$ be the center of the circular reference domain, $C$, of diameter $d$. To compute the tilts in $x-$ or $y-$direction over $C$, one minimizes the mean squared deviation, $S$, between the particular phase pattern,  $f(x,y)$, and a linear approximation: 
\begin{align} 
S  = <(a_0 + a_1\, x + a_2\, y -  f(x,y) )^2>_C
\end{align} 
$<..>$ stands for the normalized integral, i.e. the mean value, over $C$. 
\begin{align} 
dS/da_0  &= 0 =     a_0        + a_1 <x >_C +  a_2 <y >_C   -  < f (x,y) >_C \\
dS/da_1  &= 0 = a_0 <x >_C + a_1 <x^2 >_C + a_2 <xy >_C -  <x\,f (x,y) >_C \\			
dS/da_2  &= 0 = a_2 <y >_C + a_1 <xy >_C + a_2 <y^2 >_C -  <y\,f (x,y) >_C
\end{align} 
Due to the symmetry of the reference domain the moments $<x >$, $<x y >$ and  $<y >$ vanish. $<x^2 >_C =<y^2 >_C = q\,d^2$, with $q=1/16$ for a circular and $q=1/12$ for a square domain. Thus:
\begin{align} 
a_0 &=  <f >_C \\
a_1 &=  <x \, f(x,y) >_C / ( q \, d^2) \\
a_2 &=  <y \, f(x,y) >_C / ( q \, d^2) 
\end{align}

If $s$ is parallel to the direction of the slope ($a_1$), the structure function equals:
\begin{align} 
SF_x (s;d)  &=  \{ (a_1(0) - a_1(s))^2 \}    \\
&=   \{ \left(< x \, f(x,y) >_C -  < x \, f(x+s, y) >_C \right)^2  \}   / (q^2\,d^4)  \\
&=   2 \{ <  (x \, f(x,y) >_C^2 - < x\, f(x,y) >_C \, < x\, f(x+s, y) >_C \} / (q^2\,d^4)  \\
&=   2 \{ < x\, x' \,f(x,y) \,f(x',y') - x\, x' \,f(x,y)\, f(x'+s, y' ) >_C  \}  / (q^2\,d^4) \\
&=   \{ < x \, x' \, ( f(x,y) - f(x'+s, y' ))^2 - x \, x' \, (f(x,y) - f(x',y' ))^2 ) >_C  \} / (q^2\,d^4)  \\
&=   \left(\frac{\lambda}{2\pi}\right)^2  \, k\,r_0^{-5/3}\, < - x x' (u'^{5/3} - u^{5/3} )  >_C  / (2 q^2\,d^4) 	
\label{eq:SFx_z}
\end{align} 
with: $u  = ( (x - x')^2 + (y - y')^2 )^{0.5}$     and:  $u'  = ( (x - x'+ s)^2 + (y - y')^2 )^{0.5}$.

Similarly, if $s$ is perpendicular to the direction of the slope ($a_2$), the structure function equals:
\begin{align} 
SF_y (s;d)  &=   \left(\frac{\lambda}{2\pi}\right)^2  \, k\,r_0^{-5/3}\, < - y y' (u'^{5/3} - u^{5/3} )  >_C  / (2 q^2\,d^4) 
\label{eq:SFy_z}
\end{align} 

\paragraph{Approximate formulae}

\begin{figure*}
\centering
\includegraphics[height=.45\textwidth]{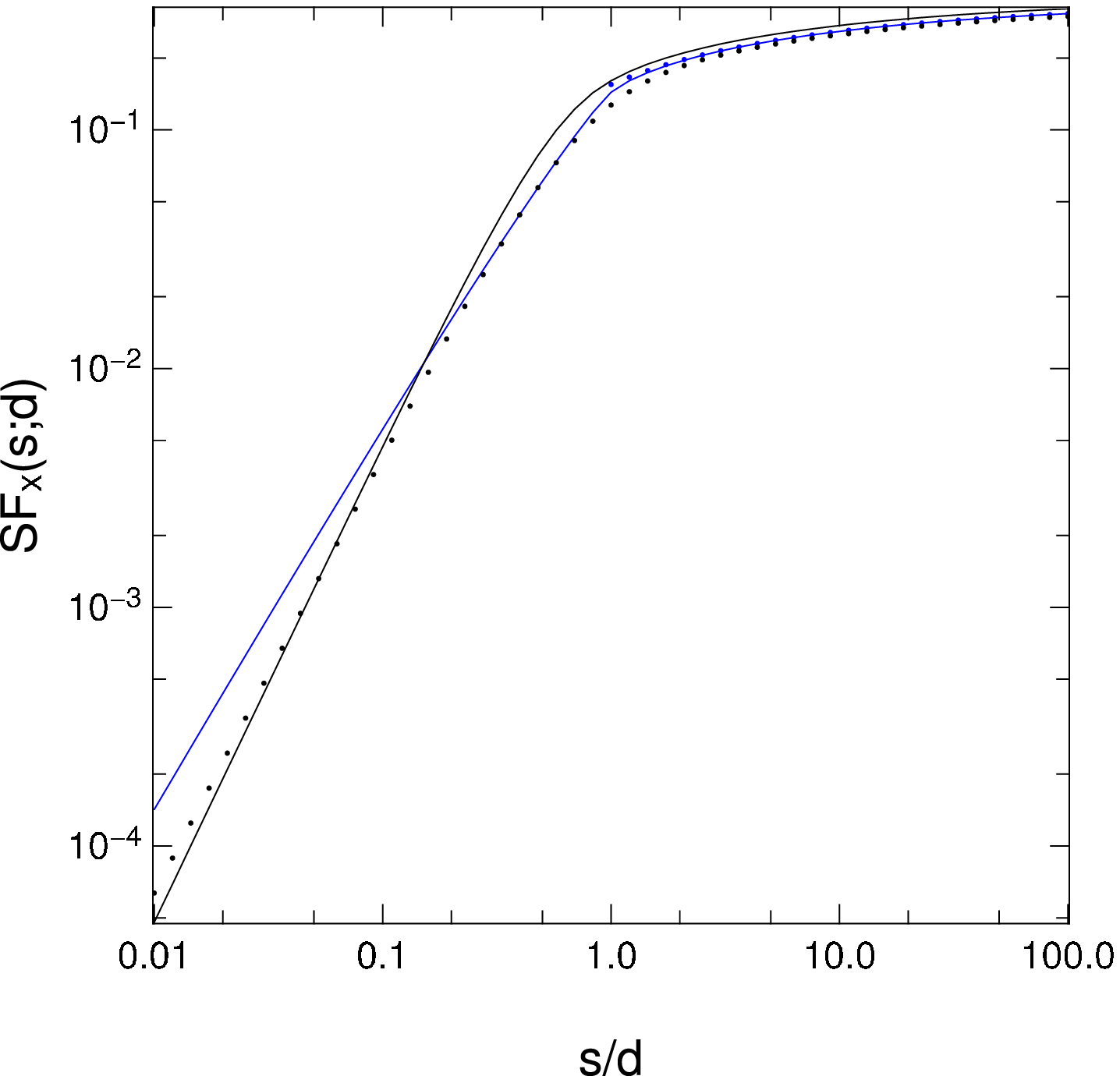}
\includegraphics[height=.45\textwidth]{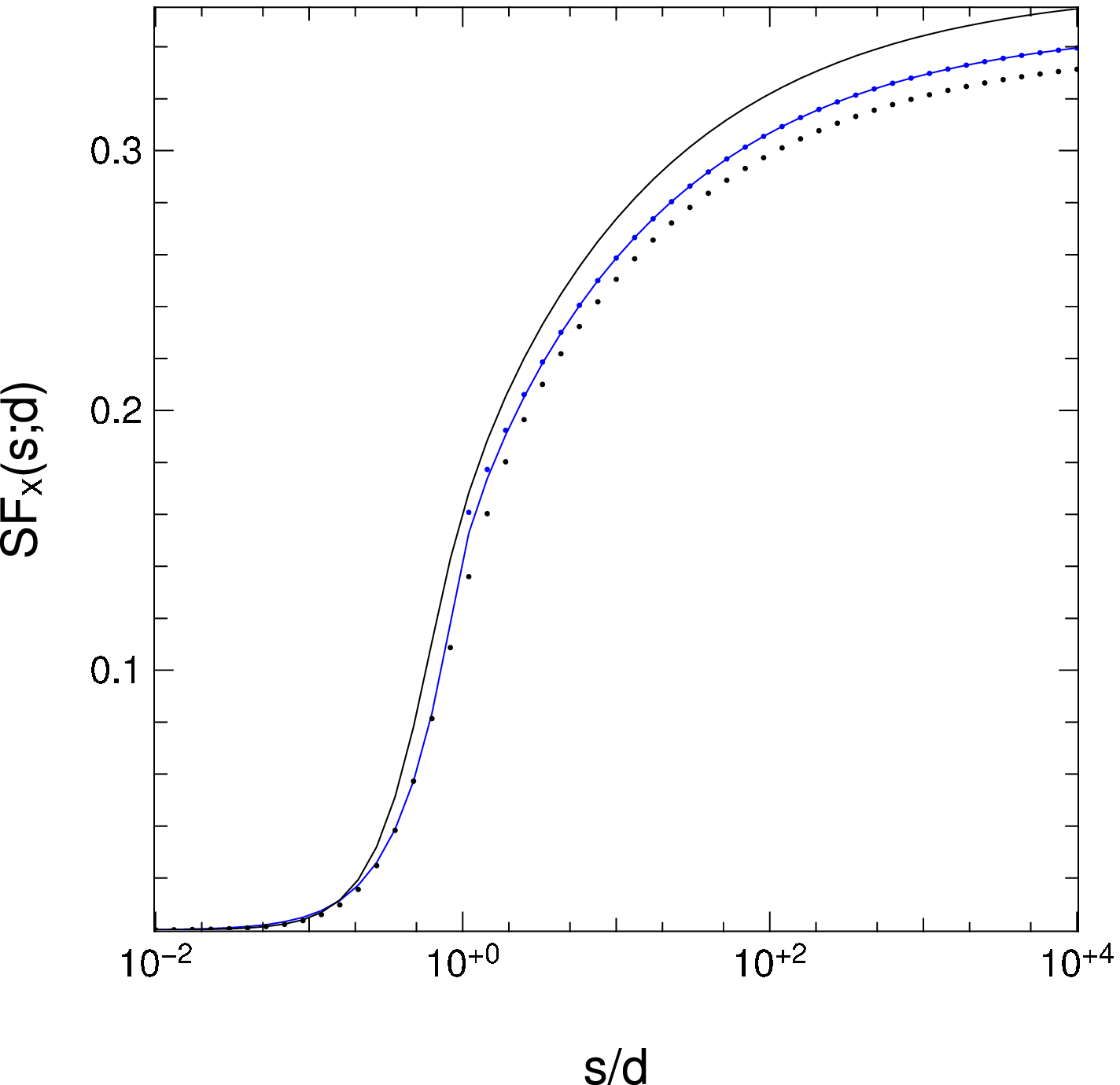}
\caption{The structure function of the $x$-slope. Solid black line: Least square slope over a circle of diameter $d$ (Z-tilt). Black dots: Average slope over a circle of diameter $d$  (G-tilt). Blue line: Average slope over a line segment of length $d$. Blue dots: Approximation used for the DIMM. Left panel: log-log scale, right panel: log-linear scale. }
\label{fig:SFx}
\end{figure*}

\begin{figure*}
\centering
\includegraphics[height=.45\textwidth]{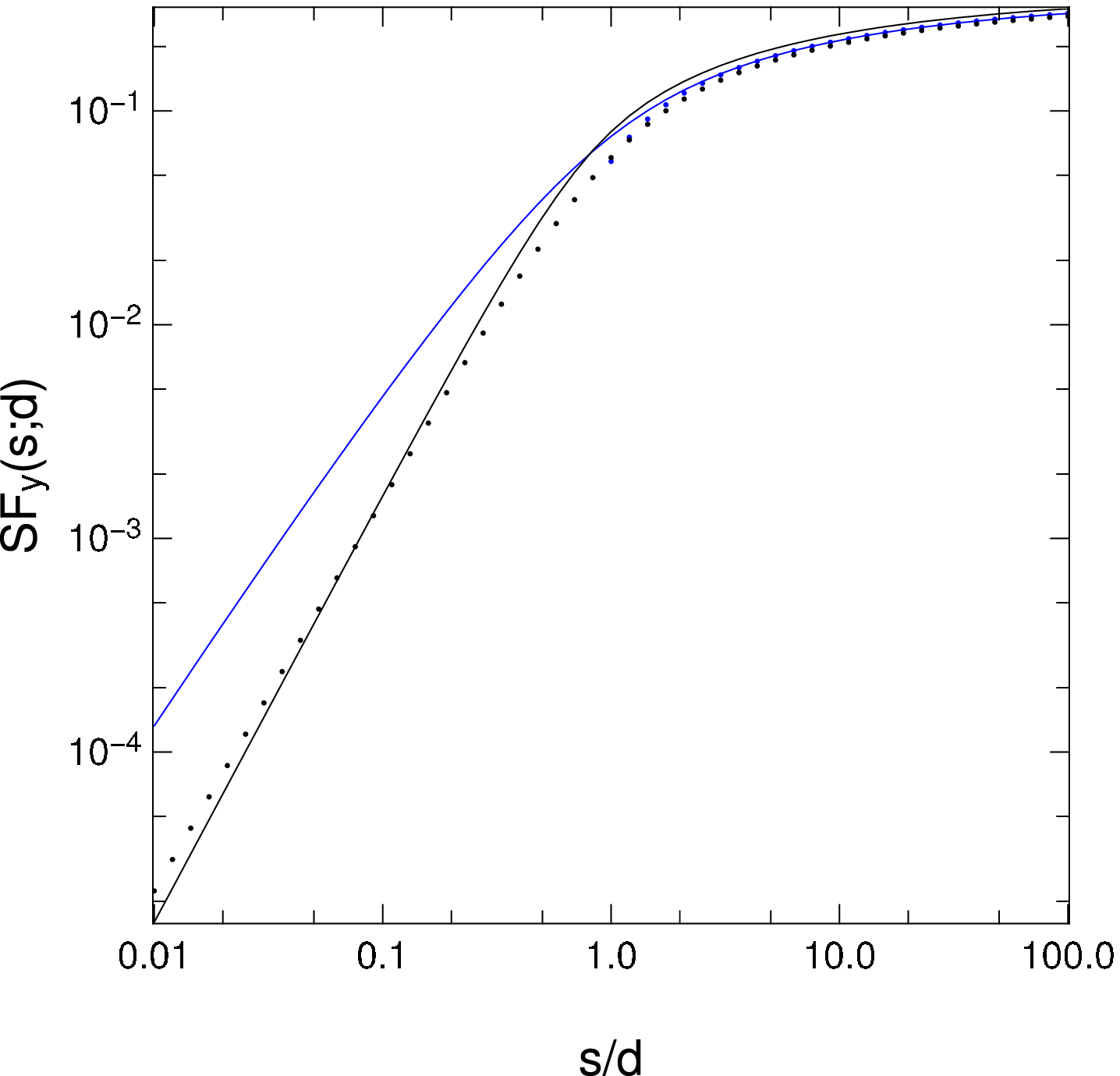}
\includegraphics[height=.45\textwidth]{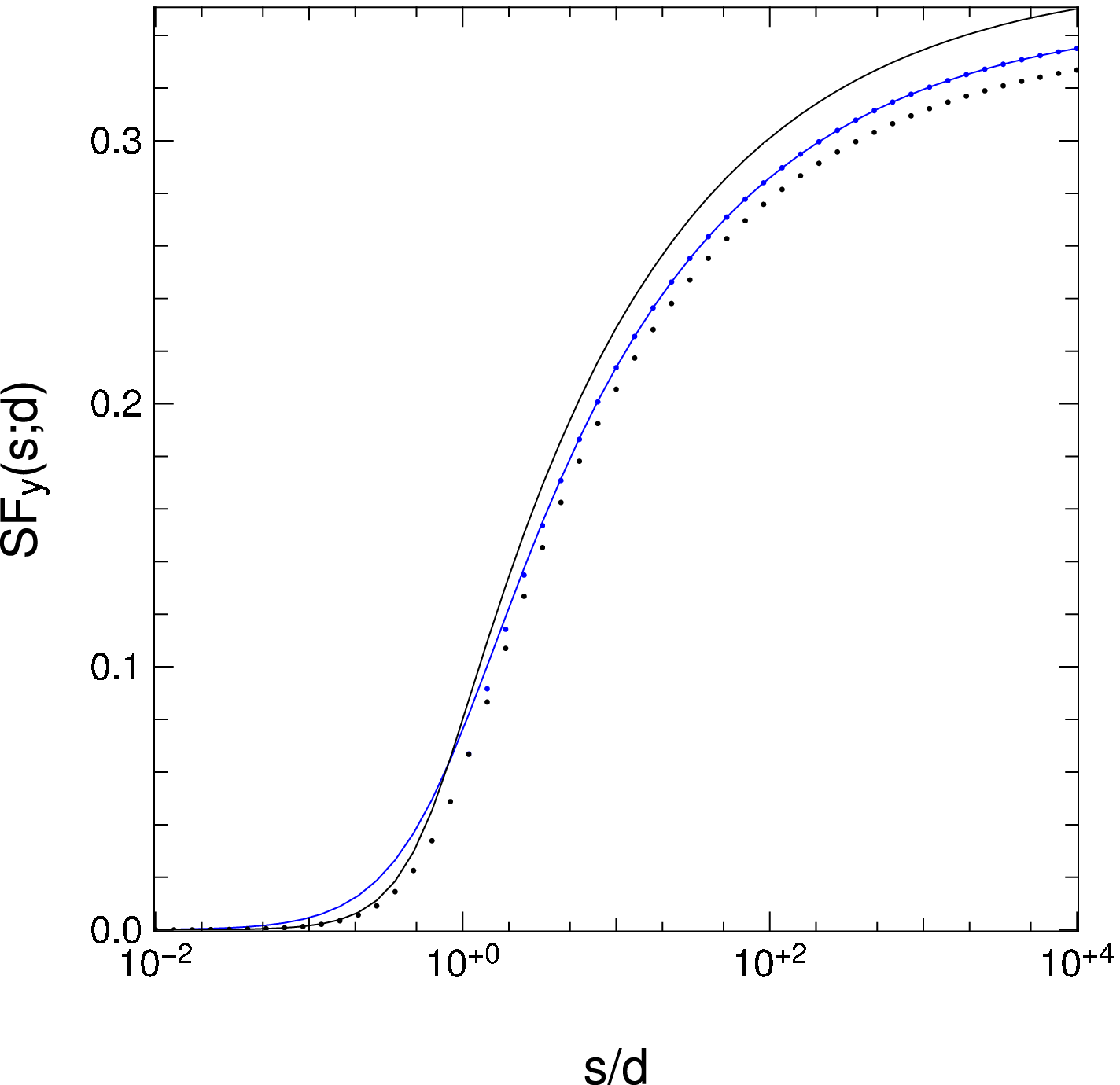}
\caption{ The structure function of the $y$-slope. Solid black line: Least square slope over a circle of diameter $d$ (Z-tilt). Black dots: Average slope over a circle of diameter $d$ (G-tilt). Blue line: Average slope over a line segment of length $d$. Blue dots: Approximation used for the DIMM. Left panel: log-log scale, right panel: log-linear scale. }
\label{fig:SFy}
\end{figure*}

\begin{figure*}
\centering
\includegraphics[height=.45\textwidth]{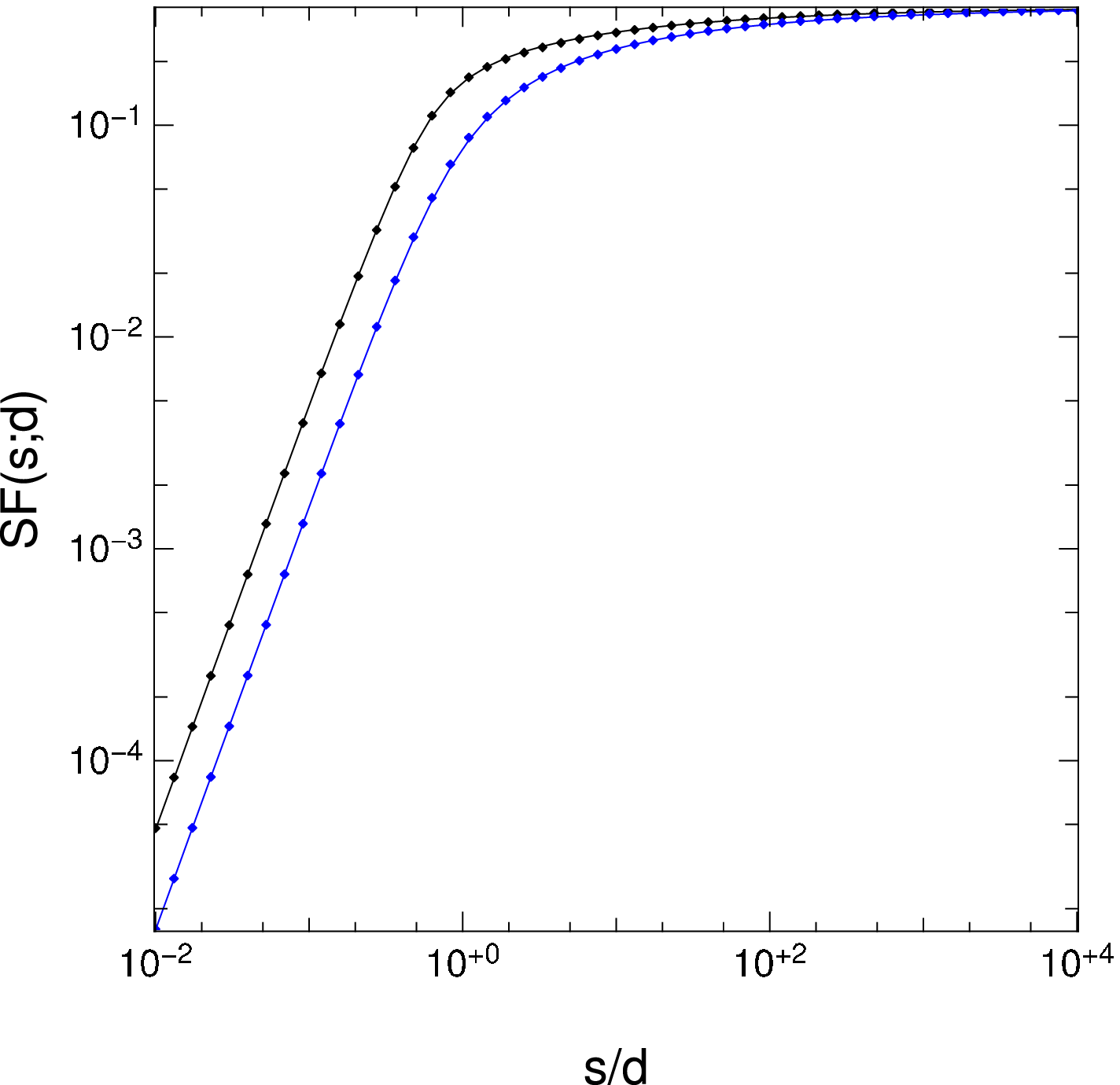}
\includegraphics[height=.45\textwidth,angle=90]{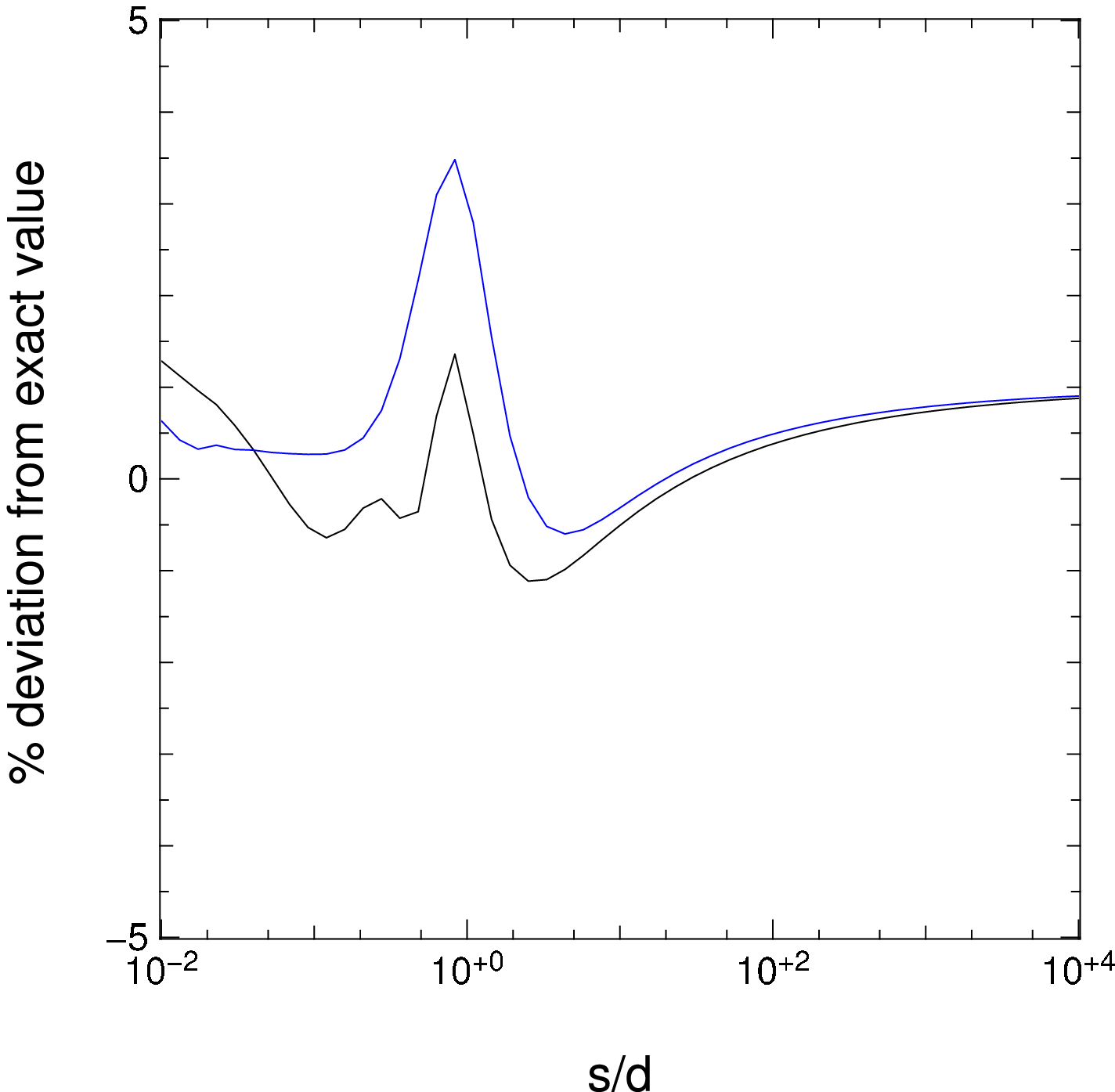}
\caption{Left panel: Structure function of tip (black) and tilt (blue) over a circle of diameter $d$. The dots show the numerical evaluation, the lines are analytical approximations (Eqs.\,\ref{eq:app_z1}--\ref{eq:app_z2}). The standard deviation is too small to to be identified. Right panel: Deviation of the analytical approximation  to the numerical values.  }
\label{fig:SF_z}
\end{figure*}

The numerical evaluation is shown on Fig.\,\ref{fig:SF_z}, it is obtained from a two-fold integration over a $60\times60$ grid. 
The following equations approximate Eqs.\,\ref{eq:SFx_z} and \ref{eq:SFy_z}, with less than 3\% deviation for $0.01<u=s/d<100$:
\begin{align} 
SF_x (s;d) &= \left(\frac{\lambda}{2\pi}\right)^2  \, k\,d^{-1/3}\,r_0^{-5/3}\, \Big(2.06-1.55\,(1+10.5\,u^2)^{-1/6} -0.51\,(1+10\,u^{3.3})^{-2/3.3}\Big) \label{eq:app_z1}\\ 
SF_y (s;d) &= \left(\frac{\lambda}{2\pi}\right)^2  \, k\,d^{-1/3}\,r_0^{-5/3}\, \Big(2.06 -1.72\,(1+1.5\,u^2)^{-1/6} -0.34\,(1+1.4\,u^2)^{-1} \Big)
\label{eq:app_z2}
\end{align} 
The numerical evaluation are represented as black lines on Figs.\,\ref{fig:SFx} and \ref{fig:SFy}.
The functions are in excellent agreement with the expressions derived by Tokovinin\,\cite{Tokovinin} (and based on calculations by Sasiela\,\cite{Sasiela}) for $s>d$: see Eq. 8 in Tokovinin\,\cite{Tokovinin}.

\section{Conclusion}

\begin{table*}[!h]
\centering
\renewcommand{\arraystretch}{1.5}
{\small
\begin{tabular}{| l p{4.5cm} | l|}  \hline
 Quantity &  & Structure function \\ \hline 
 Phase & & $k\,(s/r_0)^{5/3}\,(1+1.14\,(s/d)^{-5.5/3})^{-1/5.5}$\\ \hline
 $x$-slope & DIMM approximation & $2 \,(1-5/9\,u^{-1/3}) $\\
 & Average gradient over a line segment of length $d$ & $ 2 - | 1 - u |^{5/3} + 2 u^{5/3} - (1 + u)^{5/3} $\\
 & Average gradient over a circle of diameter $d$ (G-tilt) & $1.95 -1/9\,u^{-1/3}\,w^{-1/5.5} \,(10+1.71\,u^{-5.5/3}\,w^{-1} +8.45\,u^{-11/3}\,w^{-2} )$ \\
 & Least square slope over a circle of diameter $d$ (Z-tilt) & $2.06 -1.55\,(1+10.5\,u^2)^{-1/6} -0.51\,(1+10\,u^{3.3})^{-2/3.3}$ \\ \hline
 $y$-slope & DIMM approximation & $2 \,(1-5/6\,u^{-1/3}) $ \\
 & Average gradient over a line segment of length $d$ & $2\,( 1  + u^{5/3}  - (1+u^2)^{5/6}  ) $\\
 & Average gradient over a circle of diameter $d$ (G-tilt) & $2\,(0.98 - 1/6 \,u^{-1/3}\,w^{-1/5.5}\,(5+1.14\,u^{-5.5/3}\,w^{-1}))$\\
 & Least square slope over a circle of diameter $d$ (Z-tilt) & $2.06 -1.72\,(1+1.5\,u^2)^{-1/6} -0.34\,(1+1.4\,u^2)^{-1} $\\ \hline
\end{tabular} }
\caption{Structure functions for the degraded phase and slopes. $u=s/d$ and $w=1+1.14\,u^{-5.5/3}$. The formulae for the $x$- and $y$-slopes need to be multiplied by an additional factor $(\lambda/2\pi)^2\,k\,d^{-1/3}\,r_0^{-5/3}$.}
\centering
\label{tab:summary}
 \end{table*}

We have derived structure functions for the averaged phases and slopes from a $5/3$ power law for the structure function of the phase.
The expressions are summarized in Table\,\ref{tab:summary}, and are valid within the inertial range. 
For larger separations, the phase structure function is given by (see Eq.\,3.22 in Conan\,\cite{ConanPhD}):

 \begin{align} 
SF(s;L_0) &=  \frac{2^{1/6}\, \Gamma(11/6)}{\pi^{8/3}}\, \left( \frac{24}{5} \Gamma(6/5) \right)^{5/6} \, \left(\frac{L_0}{r_0}\right)^{5/3} \, \left(1 - \frac{2^{1/6}}{\Gamma(5/6)} \left(\frac{2\pi s}{L_0}\right)^{5/6} K_{5/6}\left(\frac{2\pi s}{L_0}\right)\right) 
\label{eq:L0}
\end{align} 

\begin{figure*}
\centering
\includegraphics[height=.45\textwidth]{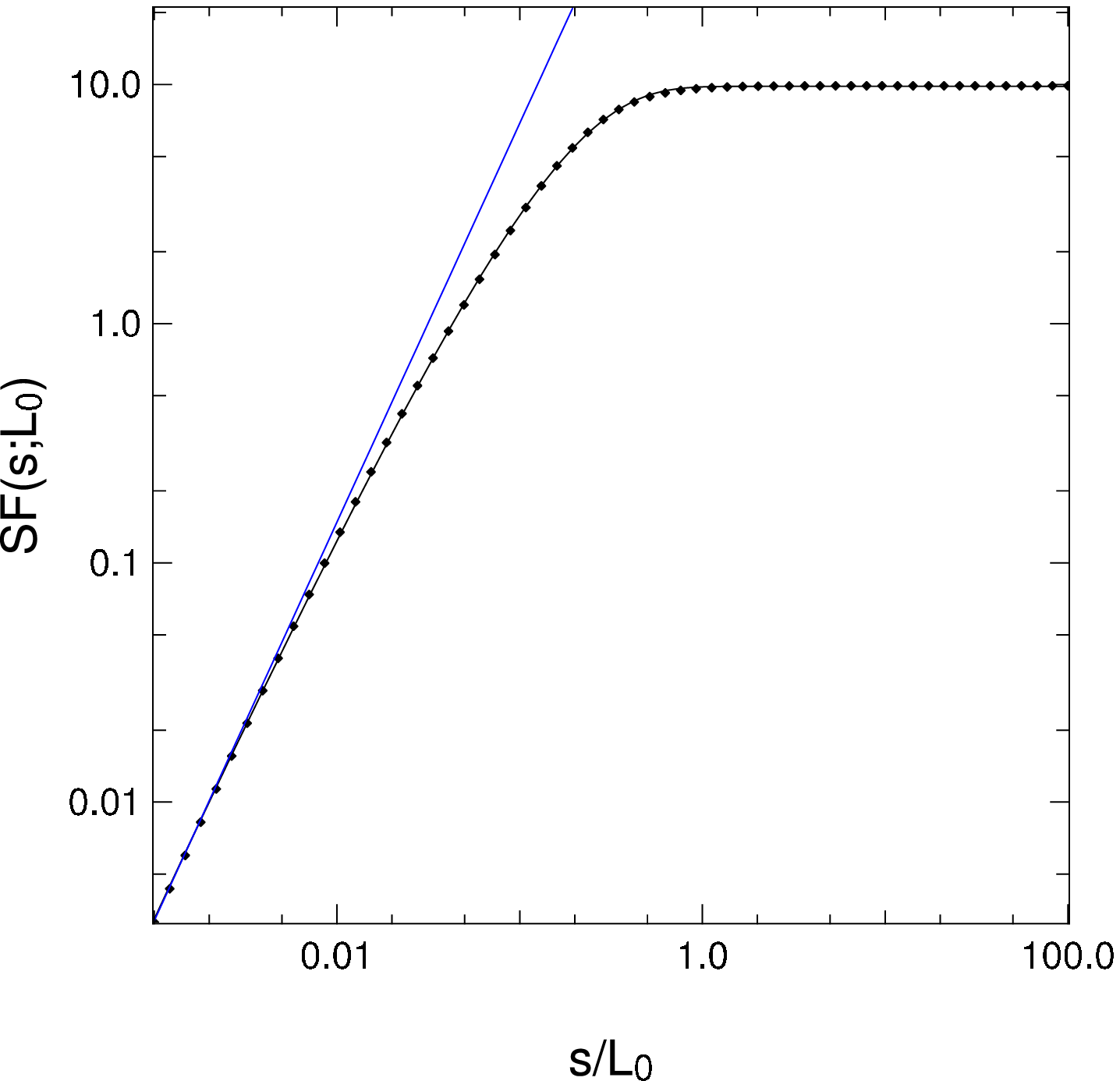}
\includegraphics[height=.45\textwidth]{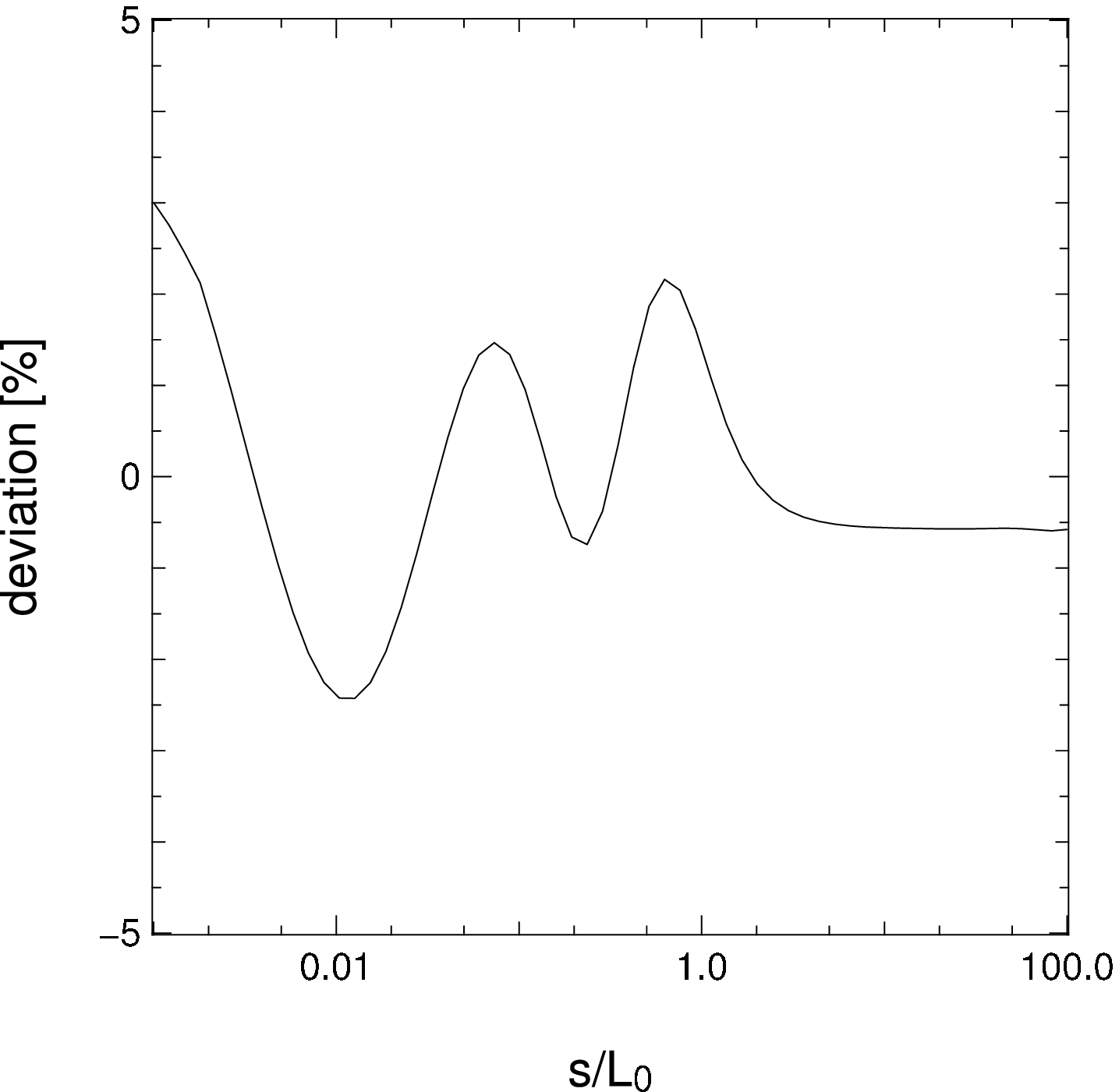}
\caption{Left panel: Structure function of the phase with finite outer scale, $L_0$. The dots represent the exact values as calculated by Conan\,\cite{ConanPhD} (Eq.\,\ref{eq:L0}), the black line shows the analytical approximation (Eq.\,\ref{eq:app_L0}). The blue line indicates the structure function with infinite outer scale (Eq.\,\ref{eq:1}). Right panel: Relative difference between the exact values and the analytical approximation.  }
\label{fig:L0}
\end{figure*}

$\Gamma$: Gamma function, $K_{5/3}$: Bessel function of the third kind and of order $5/3$.
A good approximation of this expression, is: 
 \begin{align} 
SF(s;L_0) \sim  k \,\left(\frac{L_0}{r_0}\right)^{5/3} \,\left( 1200 + 60 \left(\frac{L_0}{s}\right)^{2.3} + \left(\frac{L_0}{s}\right)^{3.4}\right)^{-5/10.2}
\label{eq:app_L0}
\end{align} 
with $k=6.88$. Fig.\,\ref{fig:L0} compares the exact expression (Eq.\,\ref{eq:L0}) with the analytical approximation (Eq.\,\ref{eq:app_L0}) and shows that the the 5/3 regime breaks off for separations $s\geq L_0/100$. 
The present approximations are thus meant to serve as useful engineering formulae when dealing with small separations, e.g. when measuring profiles of the atmospheric turbulence with site-testing telescopes such as SLODAR\,\cite{Wilson} and SDIMM+\,\cite{Scharmer}.

\appendix
\section{Shorthand notation for expectation values of phase integrals}\label{app:1}

In the text certain averages over the reference domain, $A$, are considered, that are integrals of the phase, $f(x,y)$, over $A$, or are related quantities, such as the product of the phases, $f(x,y)$ and  $f(x',y')$, of all point pairs within the region. 

To make the equations more transparent, a shorthand notation is used for the integrals. For example:
\begin{align} 
< f(x,y) \, f(x',y') >_A  =   \int_A \int_A   f (x,y) \, f (x',y') \, dx \, dy \, dx' \, dy'/ S^2
\end{align} 
	
where $S$ is the surface of $A$, and the integration runs over all point pairs $(x,y)$ and $(x',y')$ in $A$.

If the integration runs only over a function, such as $f (x,y)$, of one point, it can, of course, be written as a simple integral. But where -- in combination with other terms -- it is convenient, the double integral can nevertheless be retained, i.e. the shorthand notation can remain the same:
\begin{align} 
< f(x,y) >_A  =   \int_A \int_A   f (x,y) \, dx \, dy\, dx' \, dy'/ S^2    =   \int_A  f (x,y) \,dx \,dy / S	
\end{align} 
	
Since each wave front is given only up to a constant term, expectation values, such as $\{f(x,y)^2\}$ or $\{ f(x,y) f(x',y') \}$,  are undefined. In the equations they appear, therefore, only in combinations were the undefined terms combine to a sum of differences. In particular  $\{f(x_1,y_1)^2\}  - \{ f(x_2,y_2)^2 \} = 0$ is used to express the expectation of a sum of phase products in terms of squared differences, i.e. in terms of the structure function:
\begin{align} 
2\{ f(x_1,y_1)\, f(x_2,y_2) - f(x_3,y_3) \,f(x_4,y_4) \} &=\{ ( f(x_3,y_3) - f(x_4,y_4) )^2\} - \{(f(x_1,y_1) - f(x_2,y_2))^2 \} \\
&= SF(s_{34}) - SF(s_{12})
\end{align} 

where $s_{12}$ and $s_{34}$ are the distances between $(x_1,y_1), (x_2,y_2)$ and $(x_3,y_3), (x_4,y_4)$, respectively.

The adopted notation can be used to show the well known fact that the variance, $\sigma_f^2$, in $A$, equals half the mean squared phase difference between two points $(x,y)$ and $(x',y')$ in $A$:
\begin{align} 
\sigma_f^2 &= \{  <  f(x,y)^2 >_A  -  < f(x,y) >_A^2 \}  \\
&=\{  <  f(x,y)^2 >_A  -  < f(x,y) \, f(x',y') >_A \}  \\
&= \{   < ( f(x,y) -  f(x',y' ))^2 >_A \} / 2  \\
&=  < SF(s) >_A / 2 \\
&=  k / 2 \, < s^{5/3} >_A / r_0^{5/3}
\end{align} 
with: $s =( (x-x')^2 + (y-y')^2 )^{0.5}$. 
For a circle of diameter, $d$: 
\begin{align} 
&< s^{5/3} > = 0.3\,d^{5/3} \\
&\sigma_f^2 =  1.02\, \left(\frac{d}{r_0}\right)^{5/3}
\end{align}

\end{document}